\documentclass[%
 reprint,
 amsmath,amssymb,
 aps,
 pra,
]{revtex4-2}

\usepackage{graphicx}
\usepackage{dcolumn} 
\usepackage{bm}
\usepackage[utf8]{inputenc}
\usepackage{color}
\usepackage[hidelinks=true]{hyperref}
\usepackage{amsmath,amssymb,amsthm}
\usepackage{siunitx}

\begin{document}


\title{Mode pumping in photonic lattices using a single tailored auxiliary waveguide}

\author{David Viedma}
 \email{david.viedma@uab.cat}
\author{Jordi Mompart}%
\author{Ver\`{o}nica Ahufinger}
\affiliation{%
 Departament de F\'{i}sica, Universitat Aut\`{o}noma de Barcelona, E-08193 Bellaterra, Spain
}%




\date{\today}

\begin{abstract}
In this work, we propose a completely general method to pump the gapped topological modes of a lattice of optical waveguides by controlling the propagation constant of an auxiliary waveguide and its coupling to the main lattice. In this way, we can transform a single-waveguide excitation on the auxiliary waveguide into a specific mode of the lattice. We also demonstrate the possibility of transferring supermodes between two waveguide lattices using the same method. We illustrate the results by pumping and transferring the topological modes of Su-Schrieffer-Heeger lattices. For both scenarios, we show that purities above 99\% can be achieved with parameter values within experimental reach. Additionally, we demonstrate how the technique can be used to pump the bulk modes of the lattice for low number of waveguides or enable mode conversion between waveguide lattices.
\end{abstract}

\maketitle


\section{Introduction}

In recent years, photonic lattices with nontrivial topology have been under intense study \cite{Lu2014,Ozawa2019}. Topological states, being protected by the symmetries of the system, hold great promise for applications such as lossless transmission \cite{Mittal2014,Ma2015,Shalaev2019}, nonreciprocal processes such as unidirectional propagation \cite{Poo2011,Fang2012,Barik2018}, lasing applications \cite{Bandres2018,Ota2020}, enhancement of frequency-conversion processes \cite{Kruk2019,Wang2019}, and many others. For modes that are strongly confined, direct excitation on the waveguide with highest amplitude may prove to be sufficient \cite{Queralto2020}. However, with this method other unwanted modes may also be simultaneously excited, causing intensity beatings during the propagation. Therefore, it is of high interest to develop a general method to prepare specific modes with accuracy regardless of mode confinement, while keeping the input beam with the lowest possible complexity.

For this purpose, we propose a technique to pump topological modes of a lattice by using an auxiliary single-mode waveguide whose propagation constant is adiabatically modulated along the propagation direction. In this way, we achieve crossings between the propagation constant of the waveguide mode and that of a particular gapped supermode of the main lattice, e.g. a topological mode. The method also works for bulk supermodes of the lattice for a low number of waveguides, since they also display a gap. By controlling the coupling between the lattice and the auxiliary waveguide, the gapped supermode can be efficiently pumped, using a single-waveguide excitation on the auxiliary waveguide as the input. Even if the light transfer is not complete, the rest of the supermodes in the main lattice are not significantly excited in the process, implying a high purity of the target supermode at the output facet of the device. In a similar way, one can apply the proposed scheme to pairs of waveguide lattices. We demonstrate how modulating all waveguides in an auxiliary lattice achieves crossings between the propagation constants of its eigenmodes and those of the main lattice. This allows to transfer light between specific modes of both structures.

The technique is effectively analogous to Stark-Chirped Rapid-Adiabatic-Passage (SCRAP) between two levels. SCRAP was originally introduced as a method to transfer population between two \cite{Yatsenko1999,Rickes2000} or three \cite{Rangelov2005,Chang2007,Oberst2007,Shirkhanghah2020} atomic states, and then applied in other contexts such as wave mixing and frequency conversion \cite{Myslivets2002,Rickes2003,Wan2020,Zhang2021}. Recently, SCRAP has been adapted to waveguide optics in combination with Supersymmetry (SUSY) \cite{Viedma2021} to pump excited modes using a pair of multimode optical waveguides. In contrast, the implementation presented in this work can be used to pump any desired gapped supermode in waveguide lattices, and in particular topological modes.

Although the method is completely general, we choose the Su-Schrieffer-Heeger (SSH) model \cite{Su1979} in a lattice of single-mode optical waveguides as a platform for the study. The SSH model is the simplest instance of nontrivial topology in one dimensional (1D) lattices \cite{asboth2016}, and it can host topological edge modes localized around the ends of the chain. These states have been studied and exploited in several physical platforms \cite{Almeida2016,Bello2016,St-Jean2017,Zhao2018,Parto2018,Longhi2019AQT,Longhi2019,Muhammad2019,deLeseleuc2019,Han2020,Chaplain2020,Roy2021,Saxena2022,Viedma2022}. We showcase how we can use the proposed method to pump or transfer them between lattices with high precision. Finally, we show how the bulk supermodes of the lattice, which are completely delocalized, can be also pumped with precision using the same method, and that we can convert light into them from an arbitrary mode in the auxiliary lattice.

The present work is organized as follows: We introduce the main theoretical ideas behind the proposed pumping method in section \ref{S-Th}. After that, we present the results of the numerical simulations for the different supermodes of the SSH lattice in section \ref{S-Res}. Finally, we lay out our conclusions in section \ref{S-Con}.

\section{Theory} \label{S-Th}
\begin{figure}[t]
    \centering
    \includegraphics[width=0.9\columnwidth]{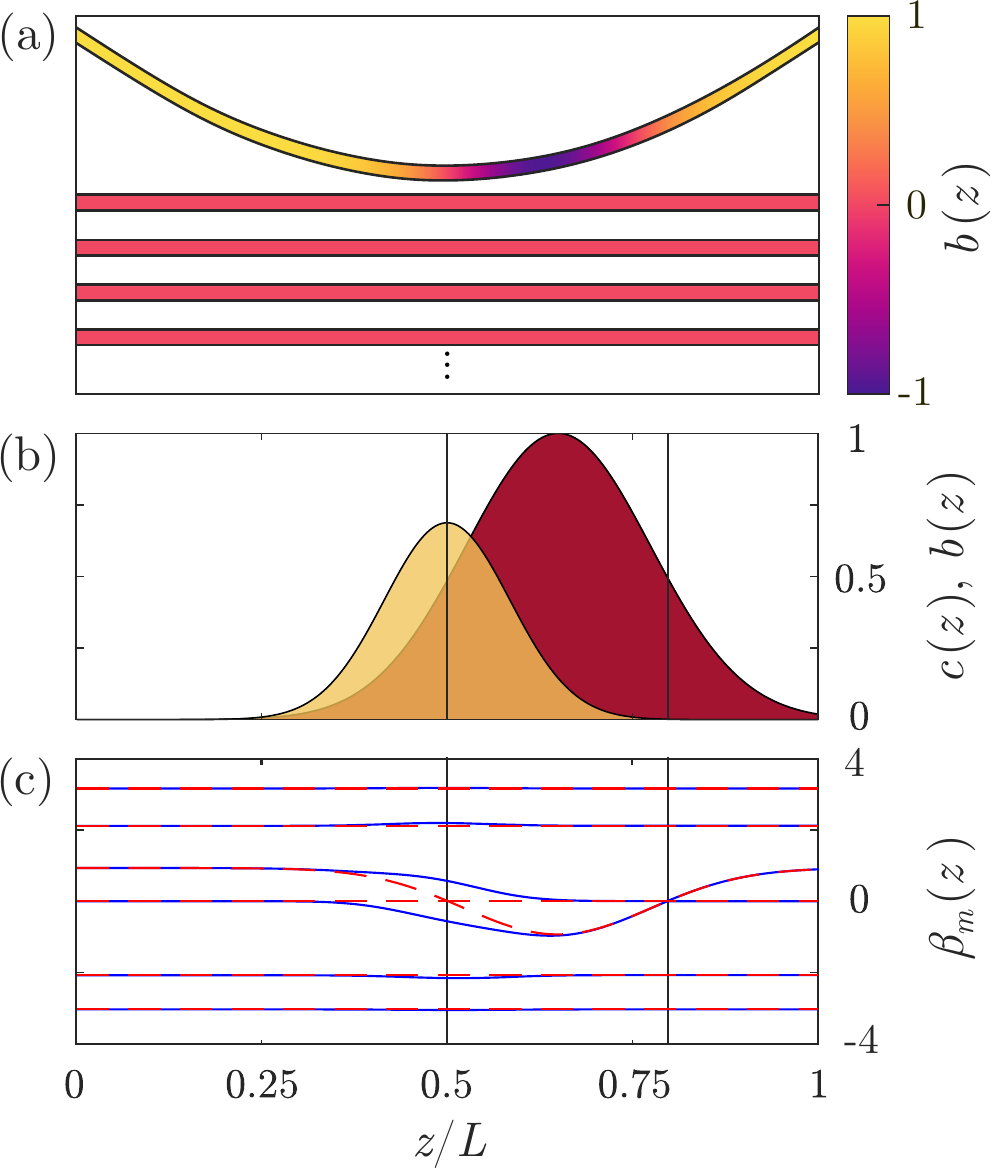}
    \caption{(a) Illustration of the geometry of the proposed device. An auxiliary single-mode waveguide has a controlled detuning (indicated by the color scheme) and controlled coupling to the main lattice along the propagation direction. The main lattice is not modulated. (b) Normalized variations of the coupling strength (yellow) and propagation constant (brown) of the auxiliary waveguide displayed in (a). (c) Mode spectrum along the propagation direction. The red dashed lines represent the propagation constants for the main lattice and the auxiliary waveguide separately, while the solid blue lines represent the ones for the modes of the joint structure. In (b) and (c), the vertical lines mark the points where the crossings occur.}
    \label{fig:1}
\end{figure}
We consider lattices of $N$ identical single-mode optical waveguides coupled to an auxiliary single-mode waveguide whose coupling and propagation constant are tailored along the propagation direction, as displayed in Fig.~\ref{fig:1}(a). We assume the tight-binding approximation to be valid, which implies that the system can be described via a coupled-mode model \cite{Jones1965}:
\begin{equation}
    i\frac{d}{dz}\bm{\psi} = \mathcal{H} \bm{\psi},
    \label{DsS-coupledmode}
\end{equation}
where $\bm{\psi} = \left(\psi_1,\ldots,\psi_N\right)^T$ are the complex modal amplitudes in each waveguide and $\mathcal{H}$ is the following tridiagonal Hamiltonian:
\begin{equation}
    \mathcal{H} = \begin{pmatrix}
        0 & c_1 & 0 & & & &\\
        c_1 & 0 & c_2 & & & &\\
        0 & c_2 & 0 &  & & &\\
         &  &  & \ddots & & &\\
         &  &  &  & 0 & c_{N-1} & 0\\
         & & & & c_{N-1} & 0 & c(z) \\
         & & & & 0 & c(z) & b(z)
    \end{pmatrix},
    \label{DsS-GeneralH}
\end{equation}
\begin{figure}[t]
    \centering
    \includegraphics[width=0.65\columnwidth]{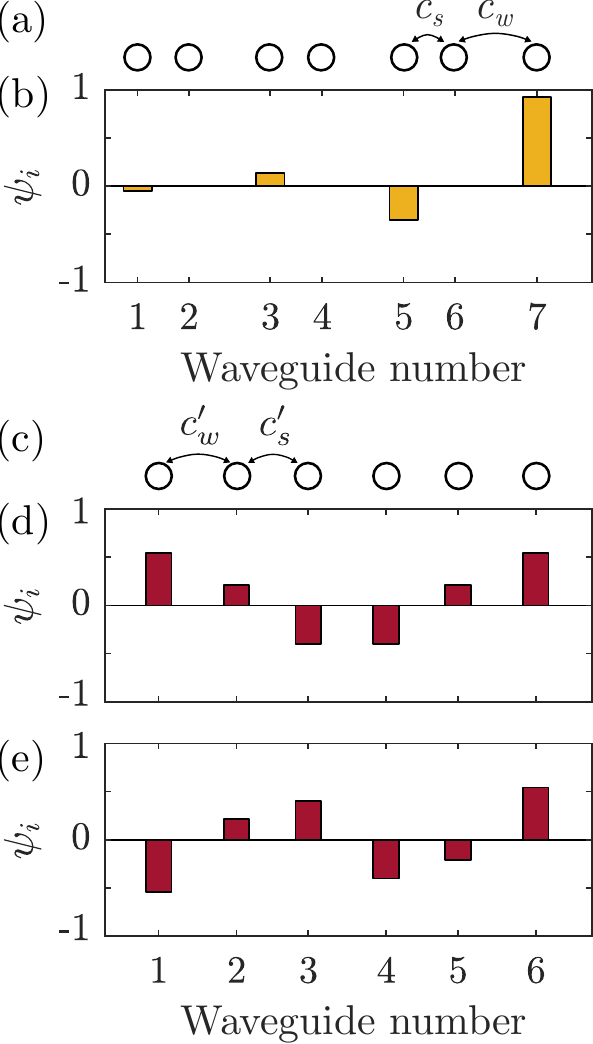}
    \caption{(a) Representation of the SSH model with odd number of sites, where the weak and strong couplings, $c_w$ and $c_s$ respectively, are represented by the different relative distance between sites. (b) Amplitudes of the edge mode hosted by the chain in (a). (c) SSH chain with even number of sites, hosting edge modes in both ends. We use $c'_w$, $c'_s$ to indicate different values from the ones in (a). (d), (e) Amplitudes of the symmetric and antisymmetric combinations of edge modes hosted by the chain in (c), respectively. For (b), $c_w = \SI{0.8}{cm^{-1}}$ and $c_s = \SI{2.1}{cm^{-1}}$; and for (d)-(e), $c'_w = \SI{2.2}{cm^{-1}}$ and $c'_s = \SI{2.5}{cm^{-1}}$.}
    \label{fig:2}
\end{figure}
where $c_i$ are the couplings between waveguides in the main lattice and where we assume that all their detunings are equal to zero. $b(z)$ and $c(z)$ correspond to the detuning of the auxiliary waveguide and its coupling to the main lattice, respectively. The proposed pumping method is based on modulating the propagation constant of the auxiliary waveguide in order to produce crossings with the propagation constant $\beta_l$ of a certain supermode of the main lattice at two points in $z$. Namely, we impose:
\begin{equation}
    b(z) = \beta_l + \Delta_0 - 2\Delta_0\exp{\left(-\frac{(z-\zeta)^2}{Z_s^2}\right)}, \label{DsS-betamodulation}
\end{equation}
\begin{figure*}[t]
    \centering
    \includegraphics[width=0.9\textwidth]{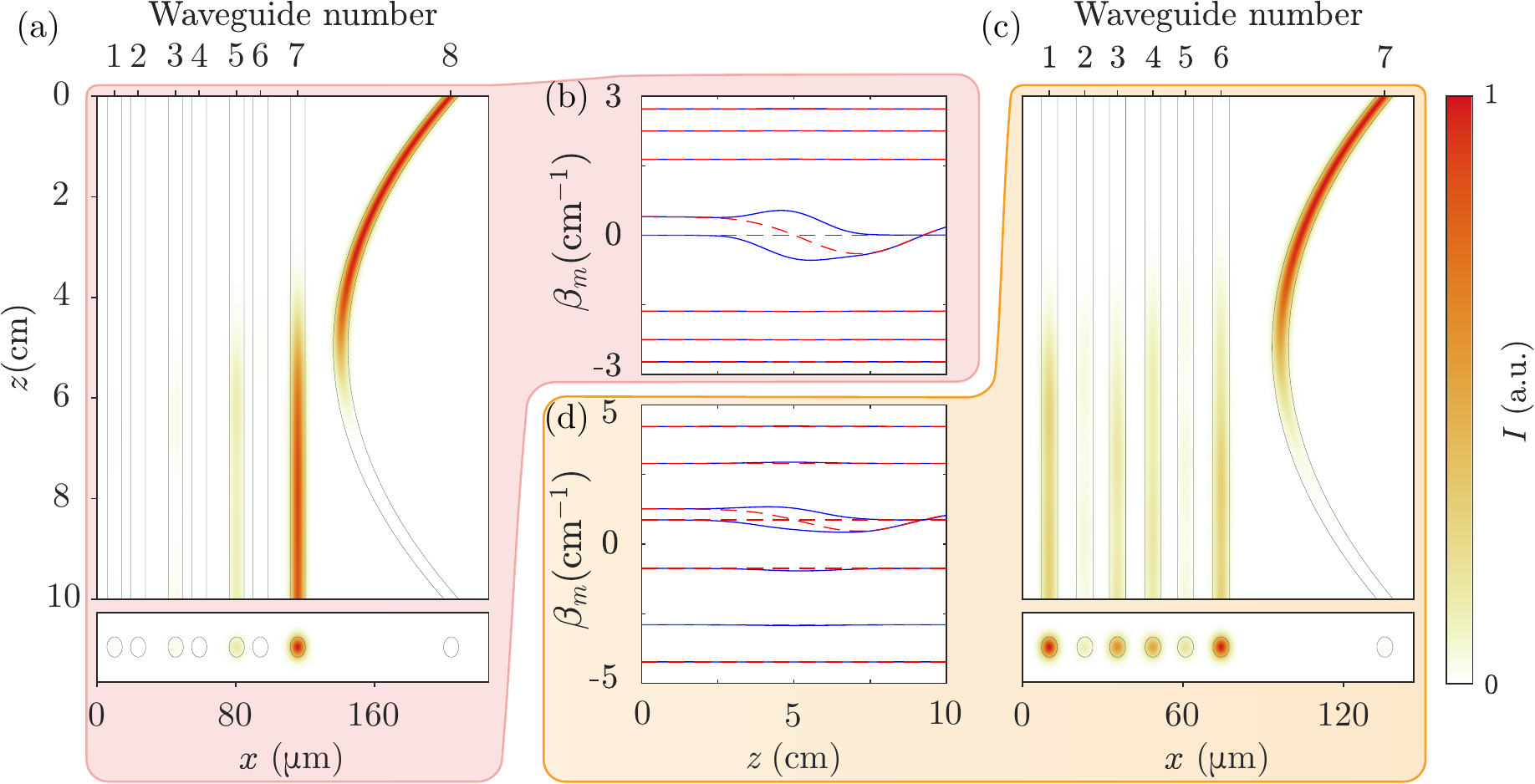}
    \caption{Light intensity propagation (a) and spectrum of propagation constants along the propagation direction (b) when pumping the single edge mode of the $N = 7$ SSH lattice with a single excitation of the auxiliary waveguide as input. We use $C = \SI{0.55}{cm^{-1}}$ and $Z_c = \SI{1.8}{cm}$. (c), (d) Same figures as in (a) and (b), respectively, for the symmetric combination of edge modes of the $N = 6$ SSH lattice with $C = \SI{0.75}{cm^{-1}}$ and $Z_c = \SI{2.3}{cm}$. For all cases, $Z_s = \SI{2.55}{cm}$ amd $\Delta_0 = \SI{0.4}{cm^{-1}}$ and the coupling parameters are the same as in Fig.~\ref{fig:2}.}
    \label{fig:3}
\end{figure*}%
where $\Delta_0 = b(0)-\beta_l$ is the initial detuning of the auxiliary waveguide with respect to the target mode $l$, and where we define $\zeta = L/2 + Z_s\sqrt{\log{2}}$, with $L$ being the length of the device and $Z_s$ the width of the modulation. To complete the pumping method, the coupling strength of the auxiliary waveguide $c(z)$ is modulated to be maximal at this first crossing and near zero during the second. By symmetrically bending the auxiliary waveguide away from the main lattice, we approximately obtain a Gaussian dependence for the coupling \cite{Queralto2017}:
\begin{equation}
    c(z) \approx C \exp{\left(-\frac{\left(z-L/2\right)^2}{Z_c^2}\right)}, \label{DsS-couplingmodulation}
\end{equation}
where $C = c_0\exp{\left(-\kappa d_{min}\right)}$ is the coupling strength at the point of minimum distance between the main lattice and the auxiliary waveguide, $d_{min}$. The parameters $\kappa$ and $c_0$ can be extracted by checking the beat length in pairs of waveguides \cite{Szameit2007}. We define $Z_c^2 = 2r/\kappa$, with $r$ being the curvature radius of the waveguide. The modulations in (\ref{DsS-betamodulation}) and (\ref{DsS-couplingmodulation}), which are sketched in Fig.~\ref{fig:1}(b), can be implemented in laser-writing setups by tuning the relative speed of the laser and the distance between waveguides, respectively \cite{Szameit2010}. Combining both modulations enables efficient light transfer between the mode of the auxiliary waveguide and the target supermode $l$ of the main lattice. For this to be true, the modulations have to fulfill the adiabaticity condition. That is, for any mode $m$, the target mode $l$ has to fulfill:
\begin{equation}
    \left|\left<\psi_l|\partial_t\psi_m\right>\right| \ll \left|\beta_m-\beta_l\right|. \label{DsS-Adiabaticity}
\end{equation}

\begin{figure*}[t]
    \centering
    \includegraphics[width=0.65\linewidth]{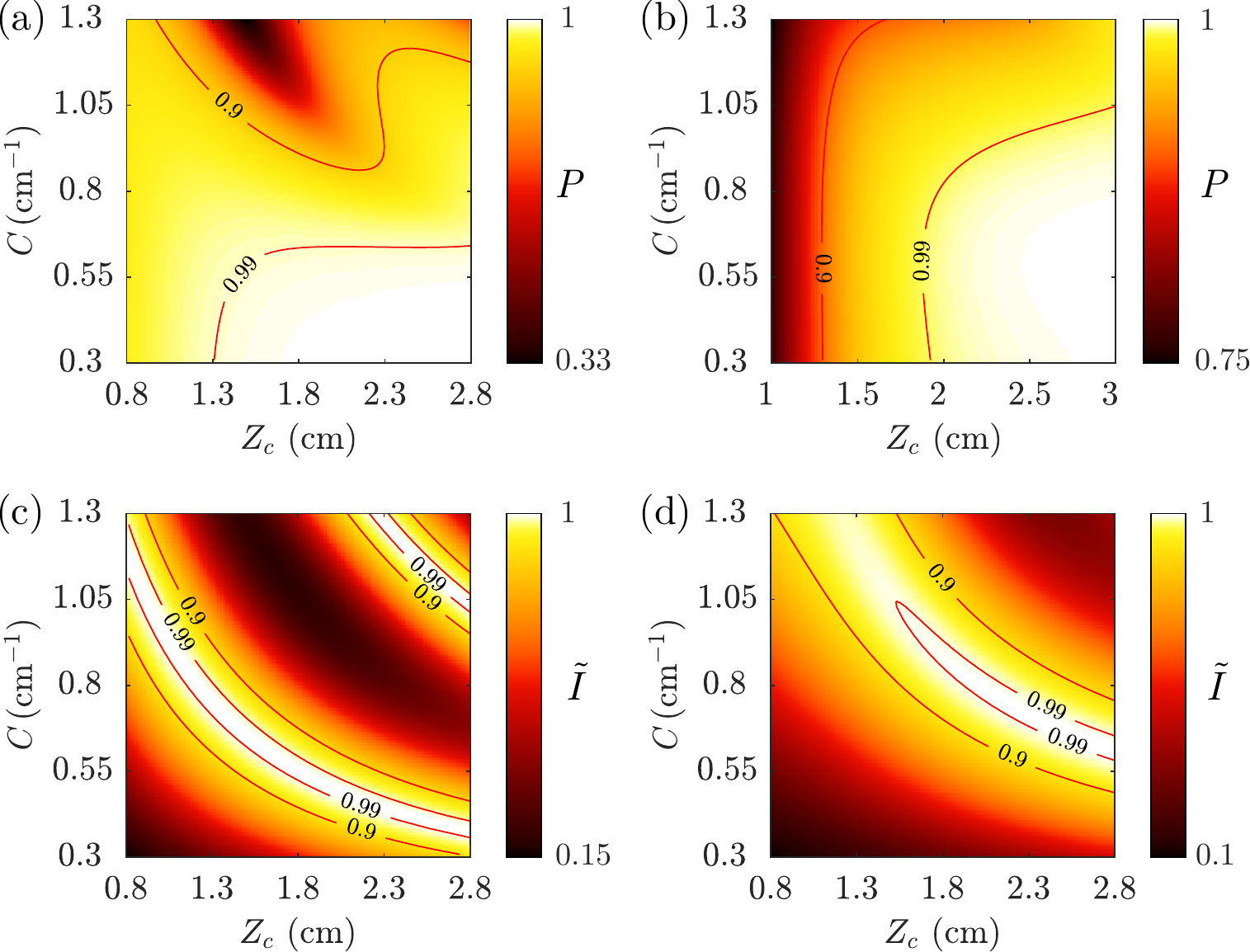}
    \caption{Purity of the output mode (top row) and intensity fraction of the pumping (bottom row) with respect to the width of the Gaussian coupling and its maximum strength for (a), (c) the edge mode of the $N=7$ lattice and (b), (d) the symmetric combination of edge modes of the $N=6$ lattice. The rest of the parameter values are indicated in Figs.~\ref{fig:2} and \ref{fig:3}.}
    \label{fig:4}
\end{figure*}
\begin{figure*}[t]
    \centering
    \includegraphics[width=0.9\textwidth]{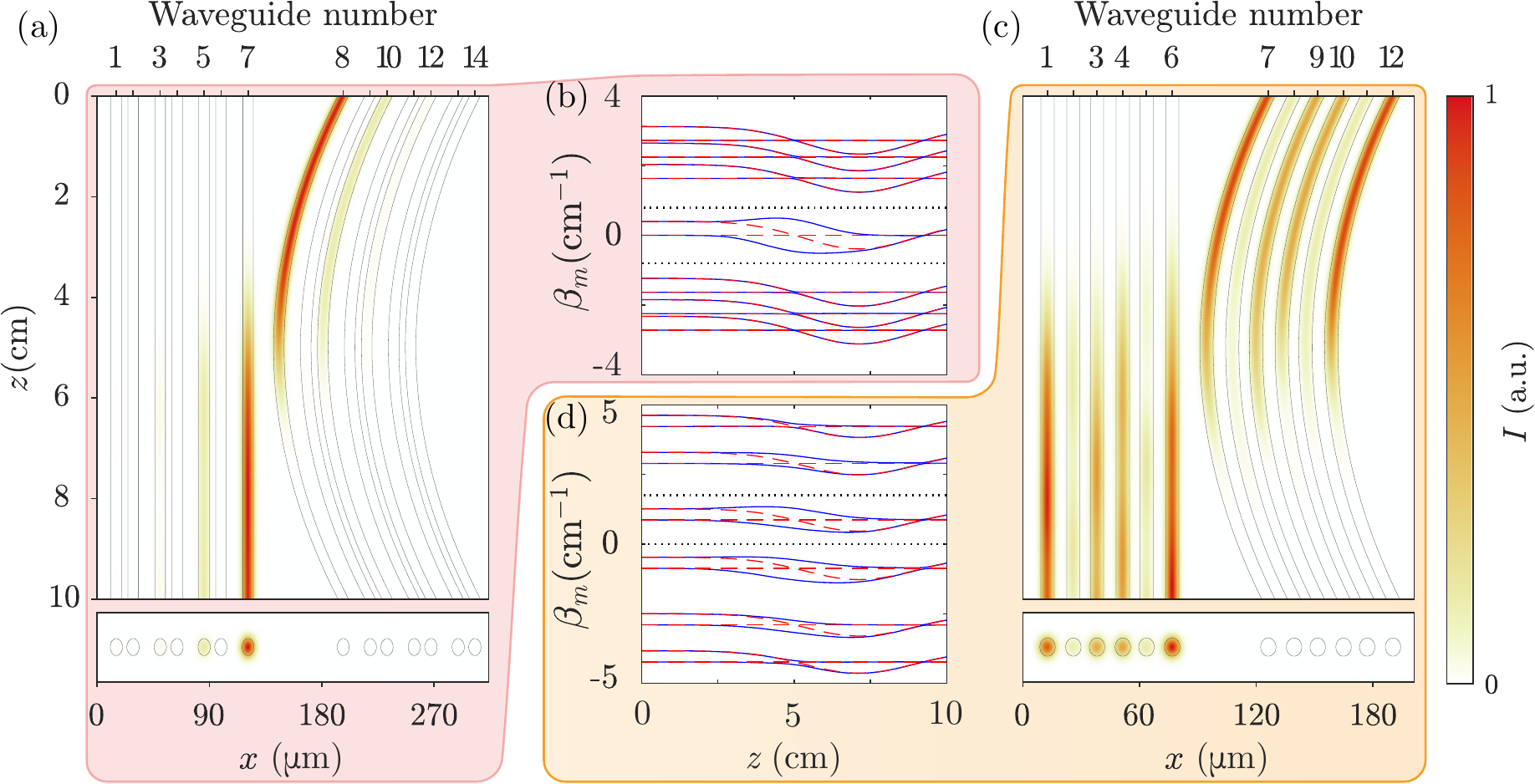}
    \caption{Light intensity propagation (a) and spectrum of propagation constants along the propagation direction (b) when pumping the single edge mode of the $N=7$ SSH lattice with the same mode of the auxiliary waveguide as input. We use $C = \SI{0.55}{cm^{-1}}$ and $Z_c = \SI{2}{cm}$. (c), (d) Same figures as in (a) and (b), respectively, for the the symmetric combination of edge modes of the $N=6$ SSH lattice with $C = \SI{1.3}{cm^{-1}}$ and $Z_c = \SI{2.55}{cm}$. The rest of the parameter values are the same as in Figs.~\ref{fig:2} and \ref{fig:3}.}
    \label{fig:5}
\end{figure*}%

The above modulations affect the spectrum of a waveguide lattice in the manner displayed in Fig.~\ref{fig:1}(c), where we use a set of equidistant lattices as an example. The crossing of the propagation constants of the mode of the auxiliary waveguide and that of a particular supermode of the lattice enables coupling between them, while the rest of the supermodes are not significantly altered. Effectively, the system can be described by a two-level Hamiltonian:
\begin{equation}
    H(z) = \begin{pmatrix}
        0 & k(z) \\
        k(z) & b(z)
    \end{pmatrix},
    \label{DsS-2w_Hamiltonian}
\end{equation}
where the coupling $k(z)$ will in general have a different value from the coupling $c(z)$ introduced in Eq.~(\ref{DsS-GeneralH}), as it will depend on the overlap between the two modes in question, but it will keep the same Gaussian dependence.

To demonstrate the potential of the method, let us now consider a lattice described by the Su-Schrieffer-Heeger (SSH) model \cite{Su1979}. The SSH model is characterized by a one dimensional (1D) lattice of identical waveguides with alternating weak and strong couplings, $c_w$ and $c_s$, respectively. The Hamiltonian describing the model is:
\begin{equation}
    \mathcal{H}_{SSH} = \begin{pmatrix}
        0 & c_w & 0 & \ldots & \ldots & 0 \\
        c_w & 0 & c_s & 0 & \ddots & \vdots \\
        0 & c_s & 0 & c_w &\ddots & \vdots \\
        \vdots & 0 & c_w & 0 & \ddots & \vdots \\
        \vdots & \ddots & \ddots & \ddots & \ddots & \vdots \\
        0 & \ldots & \ldots & \ldots & \ldots & 0
        \end{pmatrix}.
    \label{DsS-SSH_Hamiltonian}
\end{equation}
For finite lattices, the SSH model can host edge modes, which are topologically protected and exponentially localized around the edges of the lattice \cite{asboth2016}. Their propagation constants are localized at the center of the spectrum, and are gapped from the bulk bands. For the edge modes to exist, the outermost coupling of the lattice has to be the weak one, $c_w$. For an odd number of waveguides $N$, the chain will display a single edge mode at one of its ends. An example can be seen in Figs.~\ref{fig:2}(a) and \ref{fig:2}(b) for $N=7$, where we also see that these edge modes only have nonzero amplitudes in alternating waveguides. For a chain with an even number of waveguides, as the one in Fig.~\ref{fig:2}(c) with $N=6$, both ends of the lattice can display an edge mode. These will in general hybridize into symmetric and antisymmetric combinations, see Figs.~\ref{fig:2}(d) and \ref{fig:2}(e), respectively. This hybridization causes the propagation constants of these modes to split by an amount depending on the number of waveguides, $N$, and the ratio between couplings, $c_s/c_w$ \cite{asboth2016}. This ratio also controls how localized these edge modes are at the ends of the chain.

\section{Results} \label{S-Res}

\subsection{Single auxiliary waveguide}

The above formalism is now used to pump a specific topological mode of the SSH chain by tuning $b(z)$ in Eq.~(\ref{DsS-betamodulation}) to its propagation constant. We start by analyzing the SSH lattice coupled to a single modulated waveguide, in which we inject a single-mode excitation. In Figs.~\ref{fig:3}(b) and \ref{fig:3}(d), we show the variation of the propagation constants for SSH lattices of $N = 7$ and $N = 6$, respectively. In there, the red dashed lines represent the propagation constants of the supermodes of the SSH latice and of the auxiliary waveguide independently, while the blue solid lines correspond to those of the eigenmodes of the joint structure. As explained above, the dynamics is restricted to the single mode of the auxiliary waveguide, and the topological mode that we intend to pump. Particularly, for the even-$N$ chain we are restricted to small dimerizations $\left|c_s-c_w\right|$, since symmetric and antisymmetric edge modes need sufficiently different propagation constants to avoid crossing both of them during the modulation, thus ensuring that light transfers only to one of them. In Figs.~\ref{fig:3}(a) and \ref{fig:3}(c) we show the light intensity propagation through the proposed device when pumping the edge mode of a chain of odd $N$ and the symmetric combination of edges modes of a chain of even $N$, respectively. In these figures, we see that most of the light intensity is transferred into the edge modes of the main lattice while using a single excitation of the auxiliary waveguide as input.

\begin{figure*}[t]
\vspace{-5pt}
    \centering
    \includegraphics[width=0.65\linewidth]{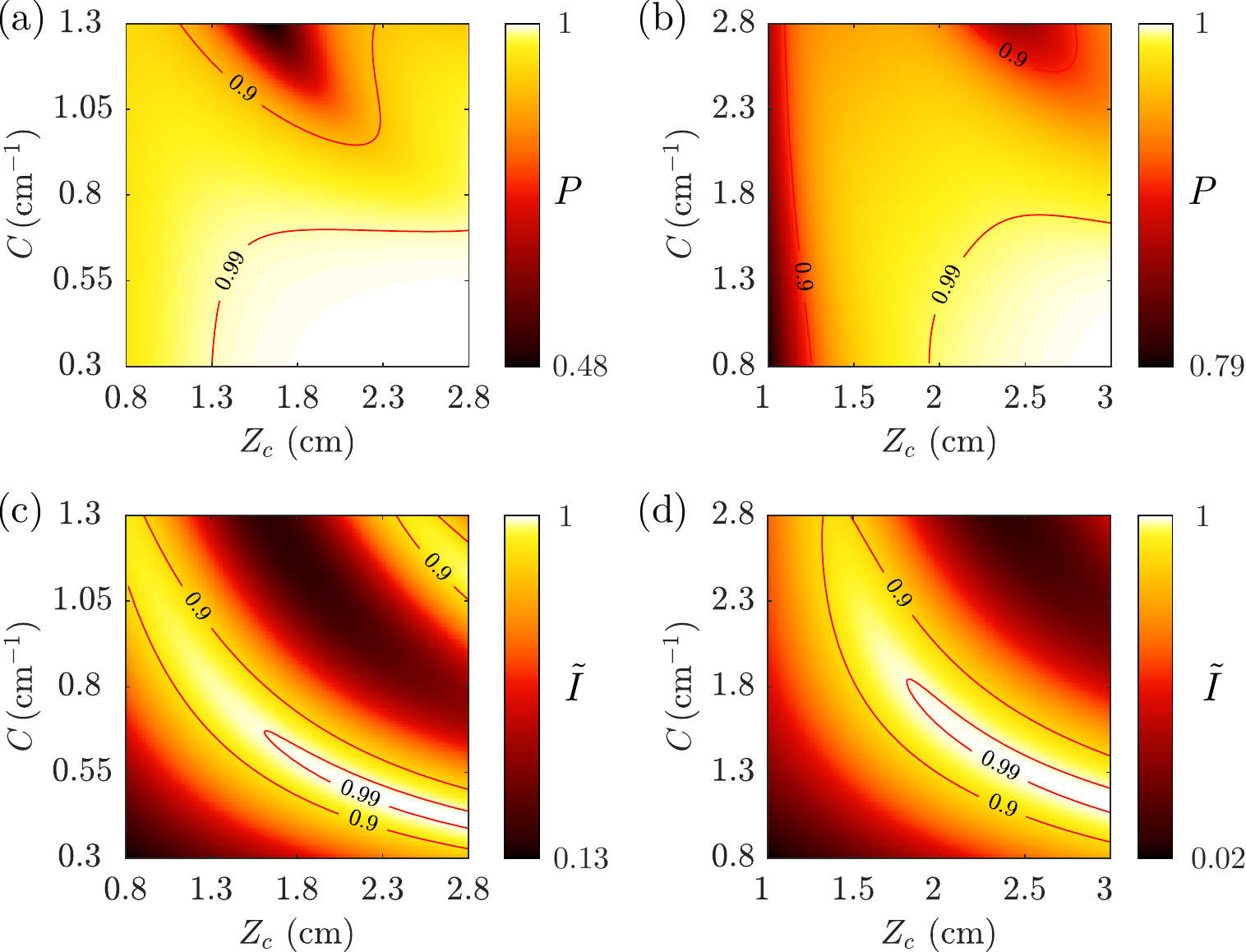}
    \caption{Purity of the output mode (top row) and intensity fraction of the transfer from the auxiliary lattice (bottom row) with respect to the width of the Gaussian coupling and its maximum strength for (a), (c) the edge mode of the $N=7$ lattice and (b), (d) the symmetric combination of edge modes of the $N=6$ lattice. The rest of the parameter values are indicated in Figs.~\ref{fig:2} and \ref{fig:3}.}
    \label{fig:6}
\end{figure*}

To check the efficiency of these devices, we define two figures of merit. First, we define the \textit{purity}:
\begin{equation}
    P=\left|\left<\bm{\tilde{\psi}}_t|\bm{\tilde{\psi}}(L)\right>\right|^2, \label{DsS-purity}
\end{equation}
which compares the normalized amplitudes of the output state $\bm{\tilde{\psi}}(L)$ to the target mode $\bm{\tilde{\psi}}_t$ in the waveguides of the main lattice. This figure of merit shows the ability to obtain a particular mode without significant excitation of any others. Additionally, we define the intensity fraction:
\begin{equation}
    \tilde{I} = \frac{I}{I_0}, \label{DsS-Intensity}
\end{equation}
which compares the intensity that is transferred from the auxiliary waveguide to the main lattice, $I$ to the input intensity $I_0$. We first plot in Figs.~\ref{fig:4}(a) and \ref{fig:4}(b) the purity of the output modes displayed in Figs.~\ref{fig:3}(a) and \ref{fig:3}(c), respectively, with respect to the parameters of the SCRAP scheme that can be controlled geometrically. Those are the maximum strength of the outer coupling $C$ between the main lattice and the auxiliary waveguide, controlled through the minimum distance $d_{min}$ between them, and the width of the Gaussian coupling function $Z_c$, controlled through the curvature radius $r$. For those figures, we keep the propagation constant modulation fixed. We see that there is a wide region of parameter values where the purity comfortably exceeds $P = 0.99$, specially for low values for the outer coupling $C$. Due to the nature of the method, which is based on the propagation constant crossing between input and target modes, it is expected that only the target mode has a significant amplitude at the output facet if the adiabatic conditions are fulfilled. The total amplitude, however, will depend on the fraction of intensity that is transferred into the main lattice. We plot this quantity in Figs.~\ref{fig:4}(c) and \ref{fig:4}(d), where we see that there exist regions in parameter space where the transfer of light is very efficient. Moreover, we see that the two devices are efficient for different sets of parameter values, in particular the case in Fig.~\ref{fig:4}(c) requires lower outer couplings than the one in Fig.~\ref{fig:4}(d) to be efficient. This is caused by the different coupling between the particular supermodes, which ultimately depends on the overlap between their spatial profiles. Choosing a region of parameter values where both $P$ and $\tilde{I}$ is large yields almost full transfer of light into the target mode. Also, even if the parameters are slightly off and we fail to transfer all light intensity from the auxiliary waveguide, only the target mode will be significantly pumped and minimal intensity will be transferred to other modes, as can be gathered from the wide parameter regions of high purity.

\begin{figure*}[t]
    \centering
    \includegraphics[width=\textwidth]{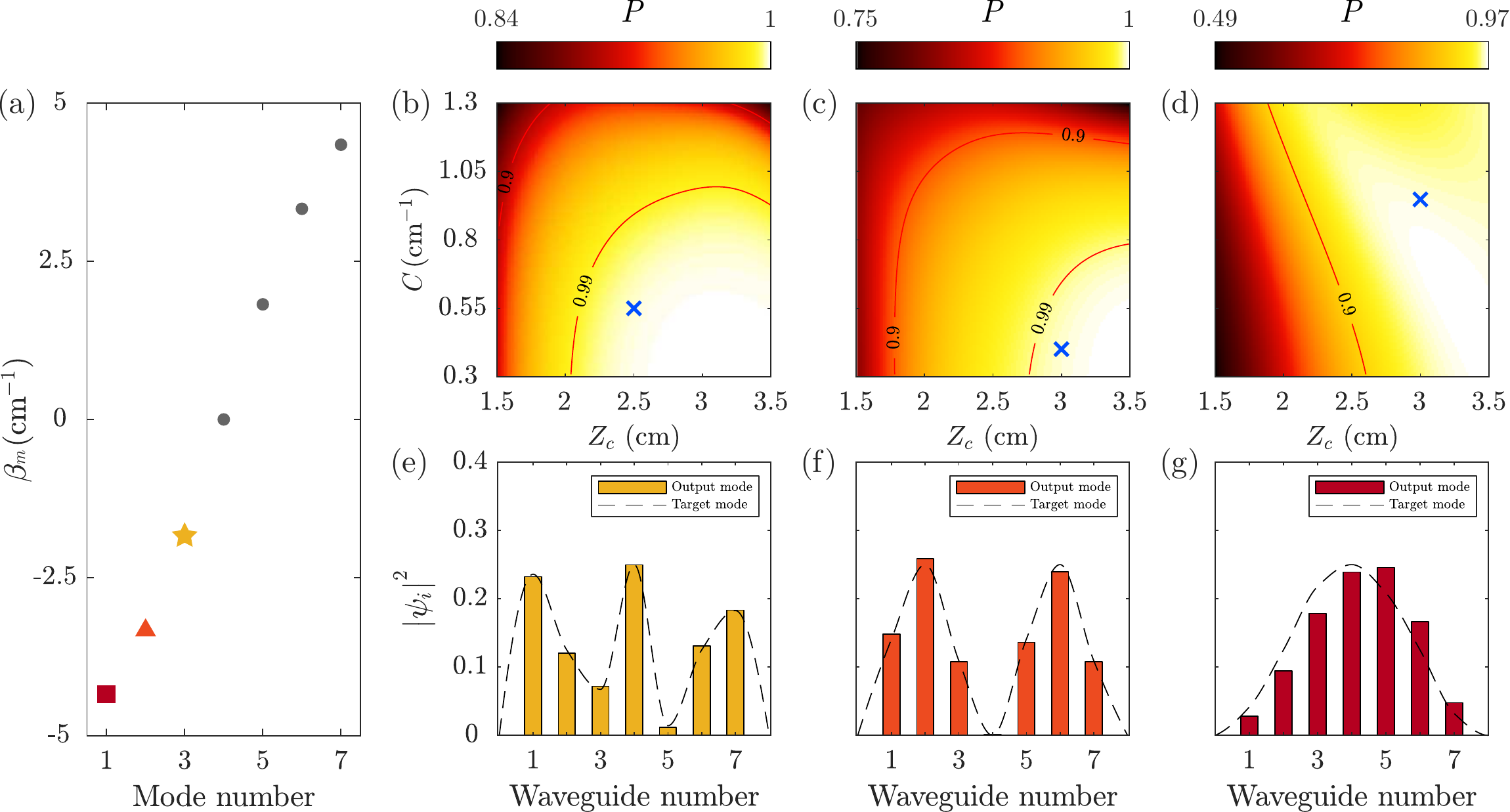}
    \caption{(a) Mode spectrum for the SSH lattice. (b), (c), (d) Purity of the output mode for the pumping of different bulk modes of the SSH lattice with $N=7$ with respect to the width of the Gaussian coupling and its maximum strength. (e), (f), (g) Output mode in the main lattice (colored bars) and target mode (dashed line) corresponding to the figures in the top row. The geometrical parameters are indicated by a blue cross in the upper figures. The propagation constants of the modes displayed in (d), (e) and (f) are marked with a yellow star, an orange triangle and a brown square, respectively, in (a). For all cases, $c_w = \SI{2.2}{cm^{-1}}$ and $c_s = \SI{2.5}{cm^{-1}}$ and the rest of the parameter values are the same as in Fig.~\ref{fig:3}.}
    \label{fig:7}
\end{figure*}

\subsection{Auxiliary lattice}

The same kind of analysis can be performed for the case of two SSH lattices, where all waveguides of the auxiliary lattice are modulated equally. In this case, the gap between edge and bulk modes also allows for efficient transfer as in the case of a single auxiliary waveguide, despite the fact that there is a larger amount of supermodes present. In fact, we produce multiple crossings since all modes of the auxiliary lattice are equivalently modulated, see Figs.~\ref{fig:5}(b) and \ref{fig:5}(d), but the dynamics is restricted to the two modes between which we want to establish the transfer. This is indicated in both Figures by the region enclosed in dotted lines. We first show in Fig.~\ref{fig:5}(a) the light transfer between an SSH lattice with odd $N$ coupled and an identical auxiliary lattice, but reflected with respect to the $x$ axis so that their respective edge states have a significant overlap. With this setup, the edge mode of the auxiliary lattice gets fully transferred into the edge mode of the main lattice. In \ref{fig:5}(c), we instead consider two lattices of even $N$. In that case, the symmetric combination of edge modes gets fully transferred between the two lattices. The modes in the auxiliary lattice could be injected using the method described in the previous section in both cases.

To compare these results with the case of a single auxiliary waveguide, we plot the purity of the output mode in Figs.~\ref{fig:6}(a) and \ref{fig:6}(b) and the fraction of transferred intensity in Figs.~\ref{fig:6}(c) and \ref{fig:6}(d). Comparing them with the respective results displayed in Fig.~\ref{fig:4}, we see that they are very similar. This is specially true for the first case of odd $N$, where the edge mode is still mostly contained within the closest waveguide of the auxiliary lattice. As such, the overlap between the modes is similar to the one in the previous section. As for the symmetric combination of edge modes, the amplitude profile in the auxiliary lattice is entirely different from that of the single-waveguide case, resulting in a lower overlap with the mode of the main lattice. Hence, the outer coupling needs to be much stronger for the light intensity to be transferred completely. This can be readily observed by comparing Figs.~\ref{fig:4}(d) and \ref{fig:6}(d). Nevertheless, we are still able to obtain purities and intensity fractions above $P=0.99$ and $\tilde{I} = 0.99$, respectively, despite the mode being delocalized throughout the whole lattice.

\begin{figure*}[t]
    \centering
    \includegraphics[width=0.7\textwidth]{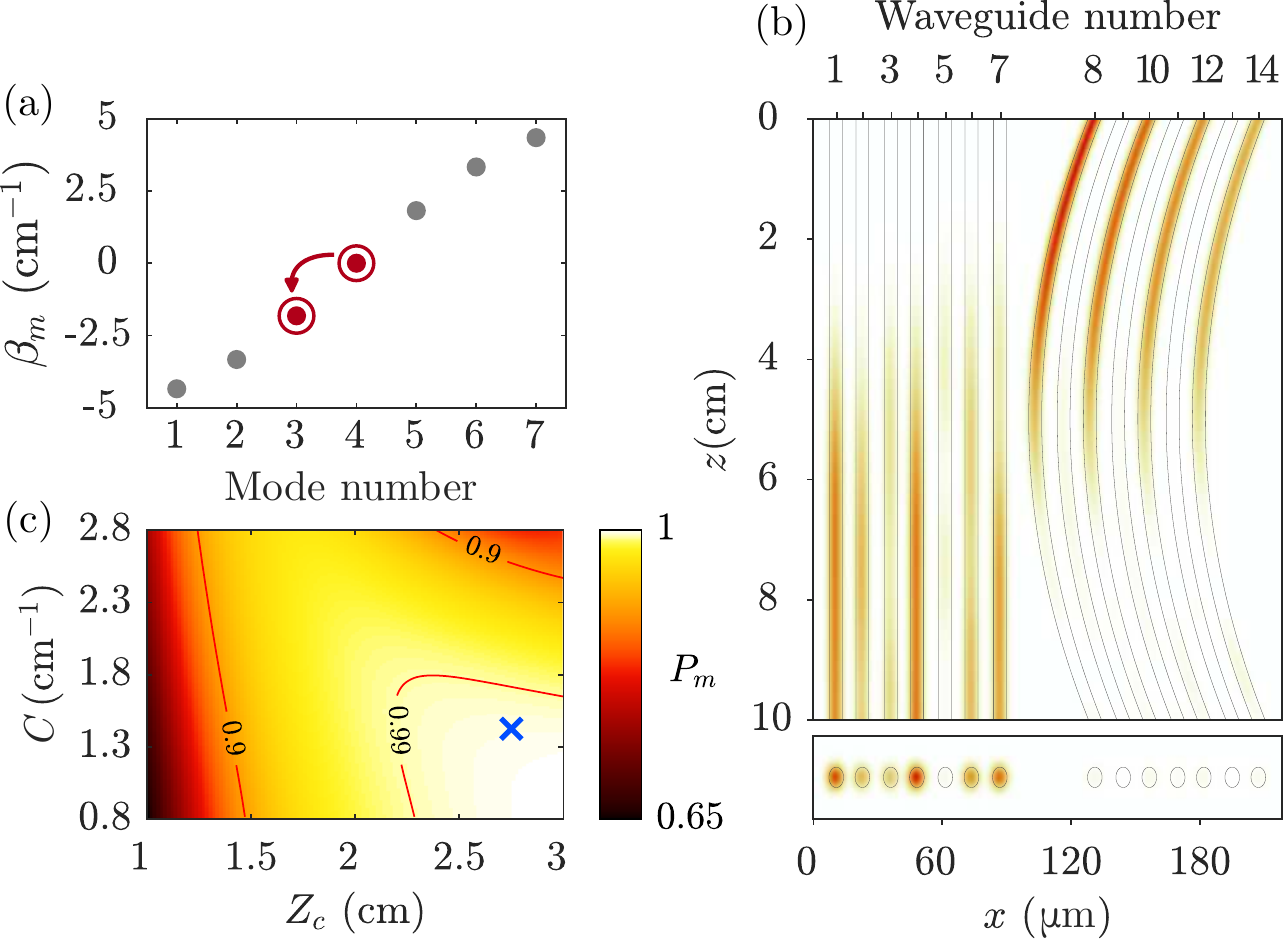}
    \caption{(a) Spectrum of propagation constants of the SSH model with $N=7$, where the input and target modes of the mode conversion are highlighted in red. (b) Light intensity propagation in the system where the edge mode of the auxiliary lattice gets transferred into a bulk mode of the main lattice. (c) Purity of the output mode in this process. The blue cross indicates the set of parameter values chosen for (a) and (b). The rest of parameter values are indicated in Figs.~\ref{fig:2} and \ref{fig:3}.}
    \label{fig:8}
\end{figure*}

\subsection{Bulk modes}

In this work, we placed the main focus on topological modes for their inherent interest. Nonetheless, the proposed pumping method goes beyond them and can be used to pump any gapped mode of the main lattice. Bulk modes are generally harder to pump, and so the fractions of transferred intensity will struggle to reach high values. However, even if the transfer is not complete and the final light intensity in the main lattice is rather low, this can be easily compensated by increasing the input beam intensity as long as the mode purity is high at the output. To showcase the potential to accurately reach these bulk modes, we consider again the case of an SSH lattice coupled to a single auxiliary waveguide. Due to the chiral symmetry of the lattice, the spectrum is symmetric around the central mode. We thus focus on one half of the spectrum, as results for the other half of the modes will be similar, we show this fact for a lattice of $N=7$ in Fig.~\ref{fig:7}(a). We tune $b(z)$ in (\ref{DsS-betamodulation}) to the propagation constant of each bulk mode in the lower half of the spectrum, and compute the purities for each of them. As shown in Figs.~\ref{fig:7}(b) through \ref{fig:7}(d), there are significant regions of parameter values where the output modes closely resembles each of the bulk modes of the lattice. This is further demonstrated by comparing the normalized output profile with the profile of the target mode in Figs.~\ref{fig:7}(e) through \ref{fig:7}(g). These results are remarkable, considering that to achieve all of them we only require a single input beam and index modulations on a single waveguide. However, the intensity fraction that can be transferred into the modes is lower than for gapped topological modes, reaching a maximum of 98\% for Fig.~\ref{fig:7}(e), 94\% for Fig.~\ref{fig:7}(f) and 61\% for Fig.~\ref{fig:7}(g).

\subsection{Mode conversion}

Another possible application of the proposed technique is enabling mode conversion between two lattices. Up until this point, when producing the modulation in the propagation constants we only considered crossings between the propagation constants of the same modes in the main and auxiliary lattices. Nevertheless, further shifting the spectrum of the auxiliary lattice can allow crossings for different modes, and as such it can lead to mode conversion between them. To prove this we consider the transfer of light from an edge mode into the bulk mode displayed in Fig.~\ref{fig:7}(e), as we sketch in Fig.~\ref{fig:8}(a). All waveguides in the auxiliary lattice are further detuned by a quantity $\Delta \beta = \beta_l - \beta_i$, corresponding to the difference in propagation constant between the initial mode $i$ and the target mode $l$, to cause crossings between their propagation constants during the modulation. Light propagation in this system is displayed in Fig.~\ref{fig:8}(b), where we see that a large fraction of light intensity, approximately 96\% in this case, gets transferred into the target mode. Checking the purity of the final mode in Fig.~\ref{fig:8}(c), we can establish that for large regions of parameter values light gets transferred into the target mode only, proving a clean mode conversion between the lattices. Note that in this case, the input edge mode is strongly delocalized due to the choice of coupling parameters for the SSH lattice.

\section{Conclusions} \label{S-Con}

In this work, we have proposed an efficient and general technique to pump gapped modes of a general lattice of coupled optical waveguides, or to transfer modes between different lattices. The technique is based on using an auxiliary waveguide -- or an auxiliary lattice -- whose relative couplings and propagation constants are controlled along the propagation direction. First, we have analyzed the performance of the technique by pumping edge modes of an SSH lattice. We have then further explored this technique for the transfer of topological modes between two SSH lattices, one being a detuned copy of the other. In both cases, purities above $P=0.99$ and intensity transfer fractions above $\tilde{I} = 0.99$ are obtained for a wide range of parameter values. Aside from topological modes, we have shown that the technique can be used to transfer light into any mode that present a gap, such as bulk modes for low number of waveguides. We have proven this both for pumping from a single waveguide excitation as well as for conversion from a different input mode in a system of two coupled lattices. 

In general, as for all adiabatic transfer methods, the transfer of light for this technique is limited by the gap between modes. The maximum change in propagation constants is limited by the presence of other supermodes of the lattice, since crossings with the propagation constants of non-desired modes can cause leaking onto them and thus spoil the transfer. Despite this limitation, we have seen that we can perform efficient transfer in waveguides with parameter values within current experimental reach. Additionally, we have seen how the purity of the output modes in the main lattice compared with the target modes remains very large even when shifting away from the regions of highest intensity transfer. Small variations in the geometrical parameters of the system, which may be caused by disorder or imperfections, mainly reduce the total light intensity at the output facet and do not cause transfer into unwanted modes, making the technique robust. Throughout the work, however, we have considered that all waveguides in the system only sustain a single mode. For an experimental implementation, one should appropriately choose the waveguide parameter values that ensure that the auxiliary waveguide remains single mode for the entire modulation of the propagation constant.

The present scheme is not limited to the cases displayed in this work, as it allows to efficiently pump or transfer light into any target mode as long as this mode displays a gap in the spectrum. The technique can be extended to other geometries exhibiting gapped topological modes or other modes of interest, such as 2D systems of optical waveguides and sets of multimode waveguides.

\begin{acknowledgments}
The authors acknowledge financial support from the Spanish State Research Agency AEI (contract no. PID2020-118153GB-I00/AEI/10.13039/501100011033) and Generalitat de Catalunya (contract no. SGR2017-1646).
\end{acknowledgments}




\bibliography{biblio}

\end{document}